\documentclass[12pt]{article}

\usepackage{epsfig}

\begin{document}

\bibliographystyle{unsrt}

\begin{flushright}
IFT 2000/2
\end{flushright}
\begin{center}
\vspace{0.6cm}
\Large{Heavy Quark Production at a Linear $e^+e^-$ and Photon Collider and 
its Sensitivity to the Gluon Content of the Photon}\\
\vspace{1cm}
\large{P. Jankowski and M. Krawczyk}\\
\vspace{0.4cm}
{\sl Institute of Theoretical Physics, Warsaw University}\\
{\sl ul. Ho\.za 69, 00-681 Warsaw, Poland}\\
\vspace{0.6cm}
\large{A. De Roeck}\\
\vspace{0.4cm}
{\sl CERN, 1211 Geneva 23, Switzerland}
\vspace{0.5cm}
\end{center}

\begin{abstract}
A  high energy linear $e^+e^-$ collider (LC) can also be used as a Photon 
Collider (PC), using Compton scattering of laser photons on the $e^+/e^-$ 
beams. The leading order  cross-section for the production of heavy quarks, 
$e^+e^- \rightarrow e^+e^-\: Q(\bar{Q}) X$, at high transverse momenta  
is calculated for both  LC and PC modes. The sensitivity of this process to 
the parton distribution parametrizations of  real photons, especially the 
gluon content, is tested for both modes.
\end{abstract}

\vspace{1.2cm}

For the study of a future electron-positron Linear Collider (LC) it is 
important to examine the physics potential for its main and possible derived 
options. The so called Photon (or Compton) Collider (PC) is an option in 
which high energy real photons can be obtained by backscattering photons from 
a laser beam on the electron or positron beam \cite{GINZ,TEL}. This way an 
excellent tool for the study of $\gamma \gamma$ collisions at high energies 
can be constructed.

In  high energy $e^+e^-$ collisions the  hadronic final state is 
predominantly produced in $\gamma^{\star}\gamma^{\star}$ interactions where 
the virtual photons are almost on mass shell. These processes can be 
described by  an effective (real) photon energy spectrum, i.e. using the
Weizs\"{a}cker-Williams (WW) approximation. A Photon Collider based on 
Compton scattering, however, provides beams of real photons, which can be 
produced in a definite polarization state and with high monochromaticity.
Moreover, the resulting photon spectrum (denoted as LASER) is much harder 
then the WW one. The comparison of the photon spectra used in this analysis
(see Appendix) is presented in Fig. 1.

The main goal of this work is to compare the LC and PC opportunities 
for probing the gluon distribution in the photon, without making use
of polarization. Heavy quark production in the unpolarized electron-positron 
scattering $e^+e^-\rightarrow e^+e^-Q (\bar Q) X$ is a promising process for 
such a study, see e.g.~\cite{DREES, CGKKKS} and also ~\cite{DGP}, where the 
related topic is considered. The measurement of this process 
is also an important test of QCD by itself. Heavy quarks can be produced in 
$\gamma \gamma$ collisions through three mechanisms. Direct (DD) production 
occurs when both photons couple directly to the $Q\bar{Q}$ pair. In single 
resolved photoproduction processes (DR) one of photons interacts via its 
partonic structure with the second photon. When both photons split into a 
flux of quarks and gluons, the process is labelled a double resolved photon 
(RR) process.

Calculations for processes involving heavy quarks are performed in two 
schemes. They differ by the number of quark flavours which are considered
to be part of the structure of the photon, and thus can take part in
the process as partons. In the case of the  massive scenario, the so called 
Fixed Number Flavour Scheme (FFNS), the photon ``consists'' only of light 
quarks and gluons, which may interact, and massive heavy quarks can be only 
created e.g. via gluon-gluon fusion. The massless Variable Number Flavour 
Scheme (VFNS) considers apart from the gluons and $u, d$ and $s$ quarks 
also heavy quarks as active flavours, which are all treated massless. This 
scheme is expected to be valid only for $p_T >> m_{c(b)}$. All the partonic 
reactions contributing in LO to heavy quark production in these schemes are 
shown in table 1.

We calculate in LO QCD the production rates for $c$ and $b$ quarks produced 
with large $p_T$ for the $e^+e^-$ colliders LEP and LC at energies of 180 
GeV and of 300, 500 and 800 GeV, respectively, and for the $\gamma \gamma$ 
PC based on the corresponding  ($e^+e^-$) LC collider. We test the 
sensitivity of considered processes to the gluonic content of the photon by 
using two different parton density parametrizations for the real photon:
GRV\cite{GRV} and SaS1d \cite{SaS}. Both these parametrizations were 
extracted from QCD fits to photon structure function data measured in 
$e\gamma$ collisions from $e^+e^-$ interactions, but have different 
assumptions for the gluon content, which is only weakly constrained by these 
measurements. The GRV and SAS1d distributions both start the evolution from 
a small starting scale, $Q^2_0 = 0.25$ GeV$^2$ and 0.36 GeV$^2$ respectively, 
a procedure which has turned out to be quite successful for the parton 
densities in the proton. Consequently both parton densities predict a rise 
of the gluon density at small $x$. The different treatment of the vector meson
valence quark distributions leads to a larger gluon component at small-$x$ 
of the photon for GRV compared to SAS1d.

In the massive (FFNS) calculations the number of active flavours ($N_f$) is 
taken to be 3. In the massless (VFNS) scheme it varies from 3 to 5 depending 
on the value of the  hard (factorization, renormalization) scale $\mu$. 
Heavy quarks are included in the computation provided that 
$\mu > m_c$ ($m_b$) with $m_c=1.6$ GeV ($m_b=4.5$ GeV) being the mass of 
$c$ ($b$) quark. When charm production is calculated in the VFNS the bottom
quarks are always excluded, hence $N_f = 3$ or $N_f = 4$. Also the QCD energy 
scale $\Lambda_{QCD}$, which appears in the one loop formula for the strong 
coupling constant $\alpha_{s}$, is affected by change of the number of active 
flavours. Therefore we denote it as $\Lambda_{QCD}^{N_f}$. We take this scale 
to be:
\begin{equation}
\Lambda_{QCD}^{3} = 232 ~~~~\Lambda_{QCD}^{4} = 200 ~~~~\Lambda_{QCD}^{5} 
= 153 \:\rm MeV
\end{equation}
as in \cite{GRV}.
If not stated otherwhise the hard scale in the calculation of the 
cross-section $\mu$ is taken to be the transverse mass of the produced heavy
quark $m_T=\sqrt{m_Q^2+p_T^2(Q)}$.

Both resolved photon contributions (DR and RR) to the process $e^+e^-
\rightarrow e^+e^-Q (\bar Q) X$  are dominated by reactions initiated by 
gluons, especially in the PC kinematic regime. This  can be seen in Fig.~2, 
where  the individual contributions to the differential LO cross-section are 
presented for the charm quark production. We study 
{\large $\frac{d^2\sigma}{dp_T^2d y}$}, 
with $y = ${$\huge \frac{1}{2}\ln \frac{E-p_L}{E+p_L} $} being the rapidity 
of the produced heavy quark and $p_T$ its transverse momentum. The results 
were obtained in the VFNS scheme  for both types of initial photon spectra. 
A fixed energy $\sqrt s$=500 GeV and $p_T$ =10 GeV for the charm quark was 
assumed. The calculation was performed using the GRV parton parametrization.
The dominance of the gluon over the quark contribution is larger for the 
PC spectrum; it is  also larger in the FFNS scheme (not shown) compared to 
the VFNS one.

Fig.~3 shows the differential cross-sections for  charm quark production
with $p_T$=10 GeV for the direct, single resolved, double resolved and total
contributions. The results were obtained in the VFNS scheme using the GRV 
parton distribution. An interesting pattern is observed. In case of
the $e^+e^-$ LC with a WW  photon spectrum  either the process is dominated 
by direct photons coupling to heavy quarks, or resolved and direct 
contributions are found to be of the same importance. The DR and RR 
contributions increase with increasing energy. Nevertheless in the range of 
the anticipated LC energies, the gluon induced reactions do not play the 
dominant role for heavy quark production. The opposite is found for a PC: 
heavy quark production is always dominated by resolved photon interactions. 
The direct contribution becomes even less important with  increasing centre 
of mass system energy. Hence, the charm production cross-section is much more 
sensitive to the parton distribution parameterization of the photon for a PC 
compared to a LC. Since for the PC option the resolved photon contributions
are clearly dominated by the processes involving gluons (Fig.~2) this option 
offers an excellent tool for measuring the gluonic content of the photon.

An important feature of the results is the observed rise of the resolved 
photon process contribution, and therefore also an increasing sensitivity 
to gluons (see below), with increasing energy. This results from the 
fact that higher energies explore regions of small Bjorken-$x_{\gamma}$
values. The minimal $x_{\gamma}$ value reached for $p_T=10$ GeV varies from 
$\sim 0.01$ for $\sqrt{s}=180$ GeV to $\sim 0.0006$ for $\sqrt{s}=800$ GeV. 
At the same time the gluon distributions differences for the GRV and SaS1d 
parametrizations are large for small $x_{\gamma}$ values. These results
are not affected by the choice of the scheme for the heavy quark calculation 
(see Fig.~4): for both the  FFNS and VFNS the cross-section is  20-30 times 
larger for the PC than for the LC.

The  sensitivity of the considered process to the gluon distribution
is studied further by comparing the predictions obtained using two different 
parton parametrizations for the photon. In Fig.~5 and 6 the ratio of the 
relative difference of cross-sections {\large $\frac{d^2\sigma}{dp_T^2d y}$} 
is presented, obtained using the GRV and SaS1d parton distribution
parametrizations in the VFNS and FFNS schemes. As expected the PC photon 
spectrum leads to a larger sensitivity than the WW spectrum for a given  
energy of the $e^+e^-$ collider: the difference between the two structure 
function parametrizations shown is 5-20\% for a WW and 25-40\% for a PC 
photon spectrum.

We presented here only the results for $c$ quark production. The 
corresponding $b$-quark production (not shown) has all the features listed 
above though the difference between the sensitivity to the gluon distribution 
at a PC compared to a  LC is smaller. All calculated cross-sections for 
beauty production are found to be even more sensitive to the gluonic content 
of the photon than the corresponding ones  for charm production, but 
the cross-sections are smaller, see also below.

Our calculation predicts a high number of the heavy quarks, $c$ and $b$,
produced at the considered centre of mass energies of the $e^+e^-$ collider 
of 300, 500 and 800 GeV for  both the LC and PC options. The event numbers 
are given in Table 2 assuming an $e^+e^-$ integrated luminosity of 100 
$fb^{-1}$, which could be achieved at a high luminosity LC with one year of 
running, and  for $p_T>$ 10 GeV. In practice, charm with a $p_T> 10 $ GeV
produced, e.g. via $D^*$ decays, can be detected in a generic LC detector 
without dedicated detectors in the rapidity range of $|y| < 1.5-2$. The charm
detection efficiency, including fragmentation fraction and branching ratios, 
is typically around a few times 10$^{-3}$~\cite{OPALF2C}. Hence the number 
of detected events from charm with such high $p_T$ will be approximately a few
thousand for the LC and several ten thousands for the PC. The latter will 
clearly allow for precision measurements of charm production and the gluon 
distribution in the photon. Some of the advantage of the PC over the LC is 
lost however due to the charm production at large $y$ in case of the PC 
(see Fig.~4), which will go undetected with the presently planned detectors.
The statistical precision of the measurements of 
{\large $\frac{d^2\sigma}{dp_T^2d y}$} will be approximately 5-10\% at the
LC and a few \% at the PC.

In conclusion, the calculated cross-sections for heavy quarks ($c$ and $b$) 
production in two photon collisions show a much higher sensitivity to the 
parton distribution parametrization of the structure of the photon in case 
of a Photon Collider compared to a $e^+e^-$  Linear Collider. This does not 
depend on the particular scheme used to calculate the heavy quark 
cross-sections. Since the resolved photon contribution is to a large extend 
dominated by gluon induced processes, especially for a high energy PC, we 
conclude that heavy quark production provides indeed a sensitive probe of 
the gluon content of the photon. Combining the above features with the much 
larger cross-sections achieved at energies of the $e^+e^-$ collisions at a PC 
favours this option for future photon structure research. A high luminosity 
$e^+e^-$ collider which drives the PC collider will however be 
essential.

\begin{flushleft}
{\bf \large Appendix}
\end{flushleft}

The simplest  Weizs\"{a}cker-Williams formula of the Equivalent Photon 
Approximation is used:
\begin{equation}
f_{\gamma}(x)=\frac{\alpha}{2\pi}(\frac{2}{x}-2+x)
\log(\frac{\mu^2}{4m_e^2})
\end{equation}
where $x = \frac{E_{\gamma}}{E_e}$, $m_e$ is the mass of the electron and 
$\mu$ is the energy scale of the process. 

In case of the Compton (LASER) mode we use the original energy spectrum of 
unpolarized photons  \cite{GINZ}:
\begin{eqnarray}
f_{\gamma}(x) = \frac{1}{\sigma_c^{np}}[\frac{1}{1-x}+1-x-4r(1-r)], 
 \nonumber \\
\sigma_c^{np} = (1-\frac{4}{\kappa}-\frac{8}{\kappa^2})\ln(\kappa+1) + 
\frac{1}{2}+\frac{8}{\kappa}
-\frac{1}{2(\kappa+1)^2}, \\
r = \frac{x}{\kappa(1-x)}, \nonumber
\end{eqnarray}
where $\kappa$ is a parameter giving the restriction of $x$ value:  
{\large $x<\frac{\kappa}{1+\kappa}$}. It is argued \cite{TEL} that the 
optimal value of $\kappa$ is 4.83, which gives a cut of $x$, 
$x_{max}=0.83$. We have chosen these values for this analysis.
Note that the part of the spectrum with $x < 0.6$ is very 
sensitive to the technical parameters of the PC such as the size of the beam.

\vspace{0.5cm}
M. Krawczyk has been partly supported by the Polish State Committee for
Scientific Research (grant 2003B 01414, 1999-2000)

\begin{table}[b]
\begin{center}
\begin{tabular}{|c|c|c|c|}
\hline
     &    DD    &    DR    &    RR    \\
\hline
FFNS & $\gamma+\gamma \to Q+\bar{Q}$ & $g+\gamma \to Q+\bar{Q}$ & $g+g \to Q+\bar{Q}$ \\
     &                               &                          & $q+\bar{q} \to Q+\bar{Q}$ \\
\hline
VFNS & $\gamma+\gamma \to Q+\bar{Q}$ & $g+\gamma \to Q+\bar{Q}$ & $g+g \to Q+\bar{Q}$ \\
     &                               & $Q(\bar{Q})+\gamma \to Q+\bar{Q}$ 
     & $q+\bar{q} \to Q+\bar{Q}$ \\
 &&  & $Q+\bar{Q} \to Q+\bar{Q}$ \\
 &&  & $Q+Q(\bar{Q}+\bar{Q}) \to Q+Q(\bar{Q}+\bar{Q})$ \\
 &&  & $Q(\bar{Q})+q(\bar{q}) \to Q(\bar{Q})+q(\bar{q})$ \\
 &&  & $Q(\bar{Q})+g \to Q(\bar{Q})+g$ \\
\hline
\end{tabular}
\end{center}
\caption{Parton reactions contributing to the process of heavy quark 
production in the  massive (FFNS) and massless (VFNS) schemes.}
\end{table}

\clearpage

\begin{table}
\begin{center}
\begin{tabular}{|c|c|c|c|c|}
\hline
                      & $c/\bar{c}$ at LC $\times 10^{-3}$& $b/\bar{b}$ at LC $\times 10^{-3}$ & $c/\bar{c}$ at PC $\times 10^{-6}$ & $b/\bar{b}$ at PC $\times 10^{-6}$ \\
\hline
 $\sqrt{s}=300\: GeV$ & 565 & 45  & 37  & 6 \\
\hline
 $\sqrt{s}=500\: GeV$ & 900 & 86 & 59 & 11 \\
\hline
 $\sqrt{s}=800\: GeV$ & 1355 & 156 & 91 & 20 \\
\hline
\end{tabular}
\end{center}
\caption{Numbers of heavy quarks with $p_T>10$ GeV produced at
a LC and a PC for an integrated $e^+e^-$ luminosity of 
$100$ fb$^{-1}$, calculated in the FFNS scheme with the GRV
parton parametrization.}
\end{table}

\begin{figure}
\vskip -5cm
\epsfig{file=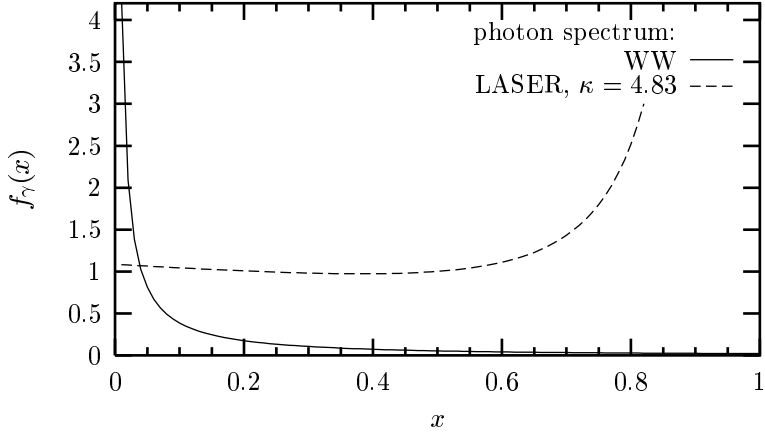}
\vskip -19cm
\caption{
Comparison of the photon spectrum $f_{\gamma}$ calculated with the
Weizs\"acker-Williams (WW) approximation, for $\mu=10$ GeV,
with the one calculated with the unpolarized laser photon spectrum (LASER),
for $\kappa=4.83$. }
\end{figure}

\begin{figure}
\vskip -5cm
\psfig{file=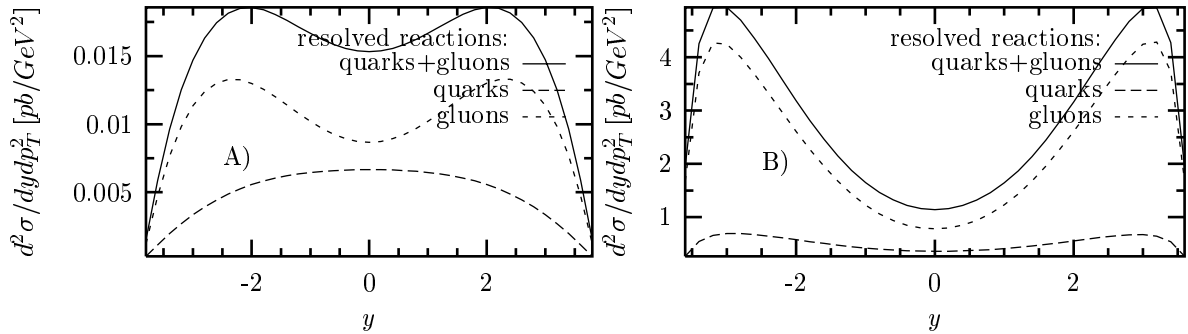}
\vskip -20.5cm
\caption{Comparison of parton reaction contributions which 
involve  gluons with
other parton reaction contributions to the resolved (DR+RR) part 
of the cross-section
{\Large $\frac{d^2\sigma}{dp^2_Tdy}$} 
$(e^+e^- \rightarrow e^+e^- c/\bar{c} X)$ 
calculated in the VFNS scheme with GRV parton parametrization at 
$\sqrt{s}=500$ GeV, $p_T=10$ GeV: A) the WW photon spectrum,
 B) the LASER photon spectrum. }
\end{figure}

\clearpage

\begin{figure}
\vskip -5cm
\psfig{file=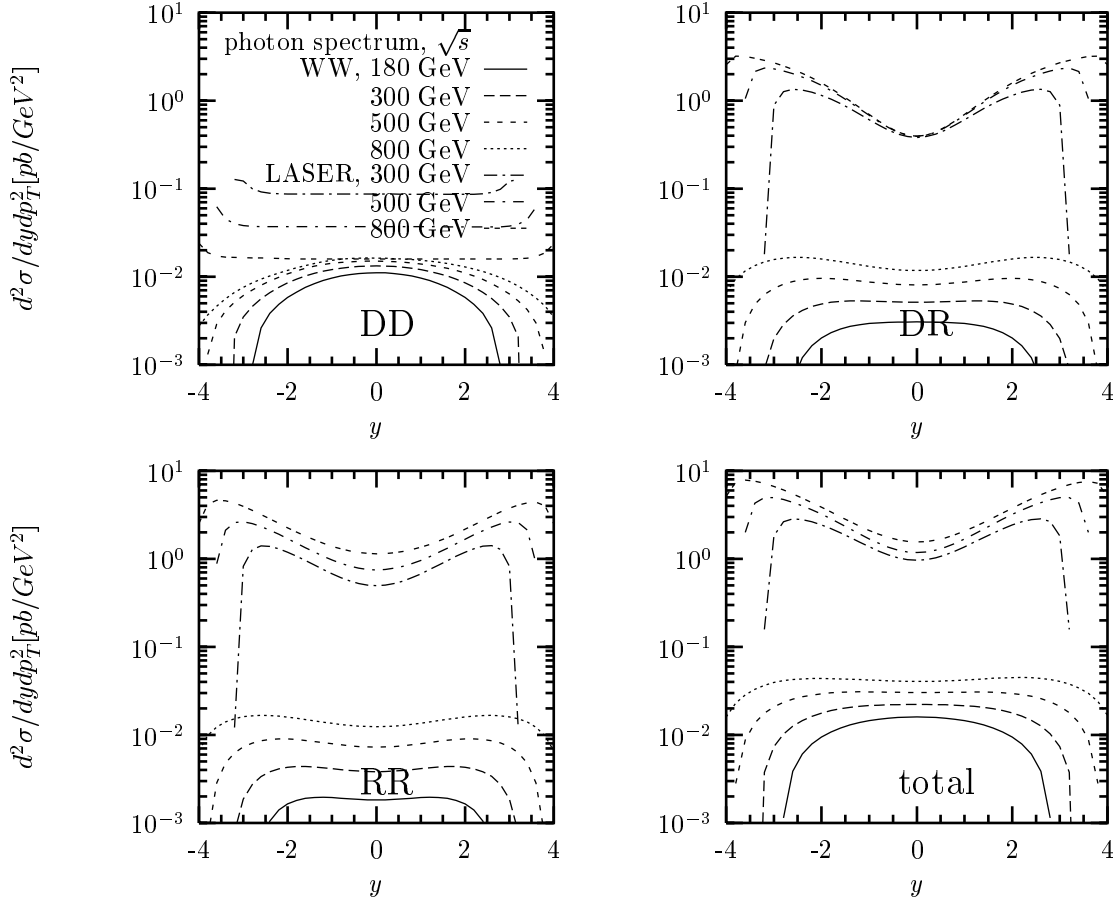}
\vskip -13.7cm
\caption{The cross-section {\Large $\frac{d^2\sigma}{dp^2_Tdy}$}
$(e^+e^- \rightarrow e^+e^- c/\bar{c} X)$ at $p_T=10$ GeV obtained with 
the GRV
parton distribution parametrization in the VFNS scheme.} 
\end{figure}

\begin{figure}
\vskip -4.5cm
\psfig{file=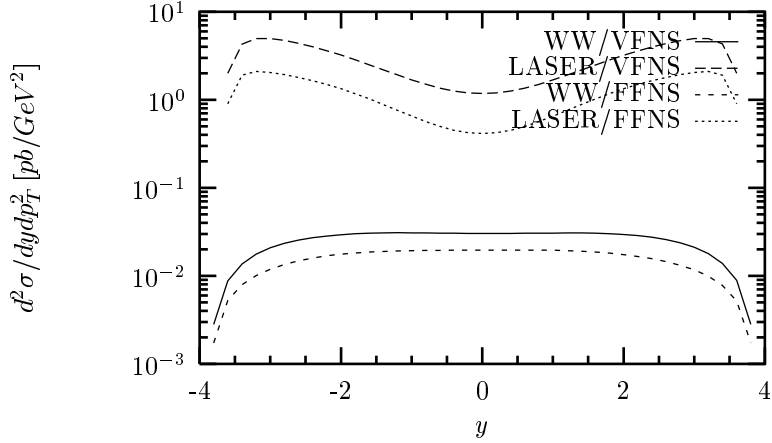}
\vskip -19cm
\caption{
The total cross-section {\Large $\frac{d^2\sigma}{dp^2_Tdy}$}
$(e^+e^- \rightarrow e^+e^- c/\bar{c} X)$ with $p_T=10$ GeV and 
$\sqrt{s}=500$ GeV calculated with the 
GRV parton distribution parametrization. Comparison 
between VFNS and FFNS schemes, and between the LASER and WW photon spectra.}
\end{figure}

\clearpage

\begin{figure}
\vskip -5cm
\psfig{file=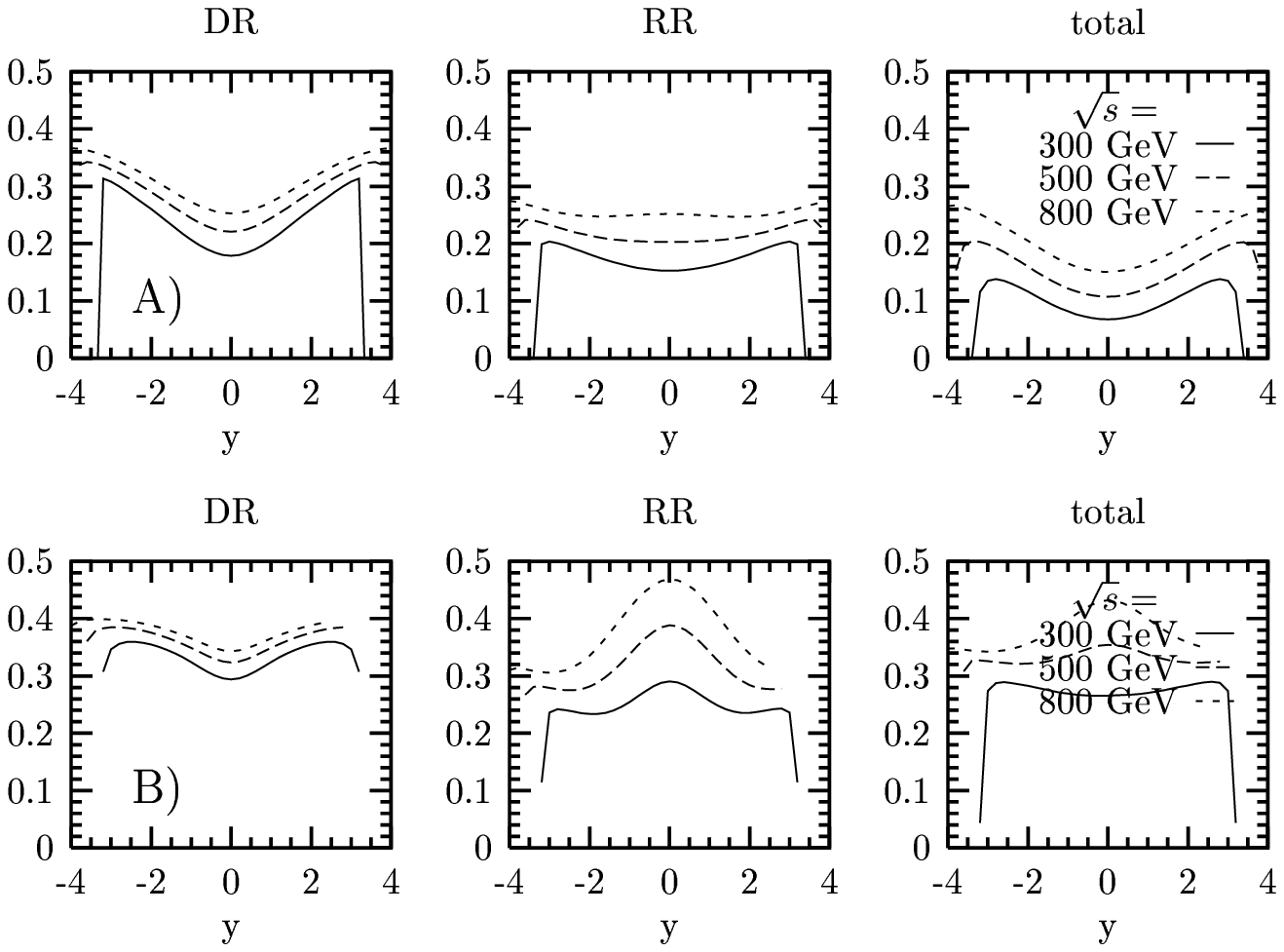}
\vskip -17cm
\caption{ The ratio {\large $\frac{GRV-SAS1d}{GRV}$} 
of the cross-section {\Large $\frac{d^2\sigma}{dp^2_Tdy}$}
$(e^+e^- \rightarrow e^+e^- c/\bar{c} X)$
\newline for $p_T=10$ GeV, in VFNS scheme, for two photon spectra: 
A) WW, B) LASER }
\end{figure}


\begin{figure}
\vskip -4cm
\psfig{file=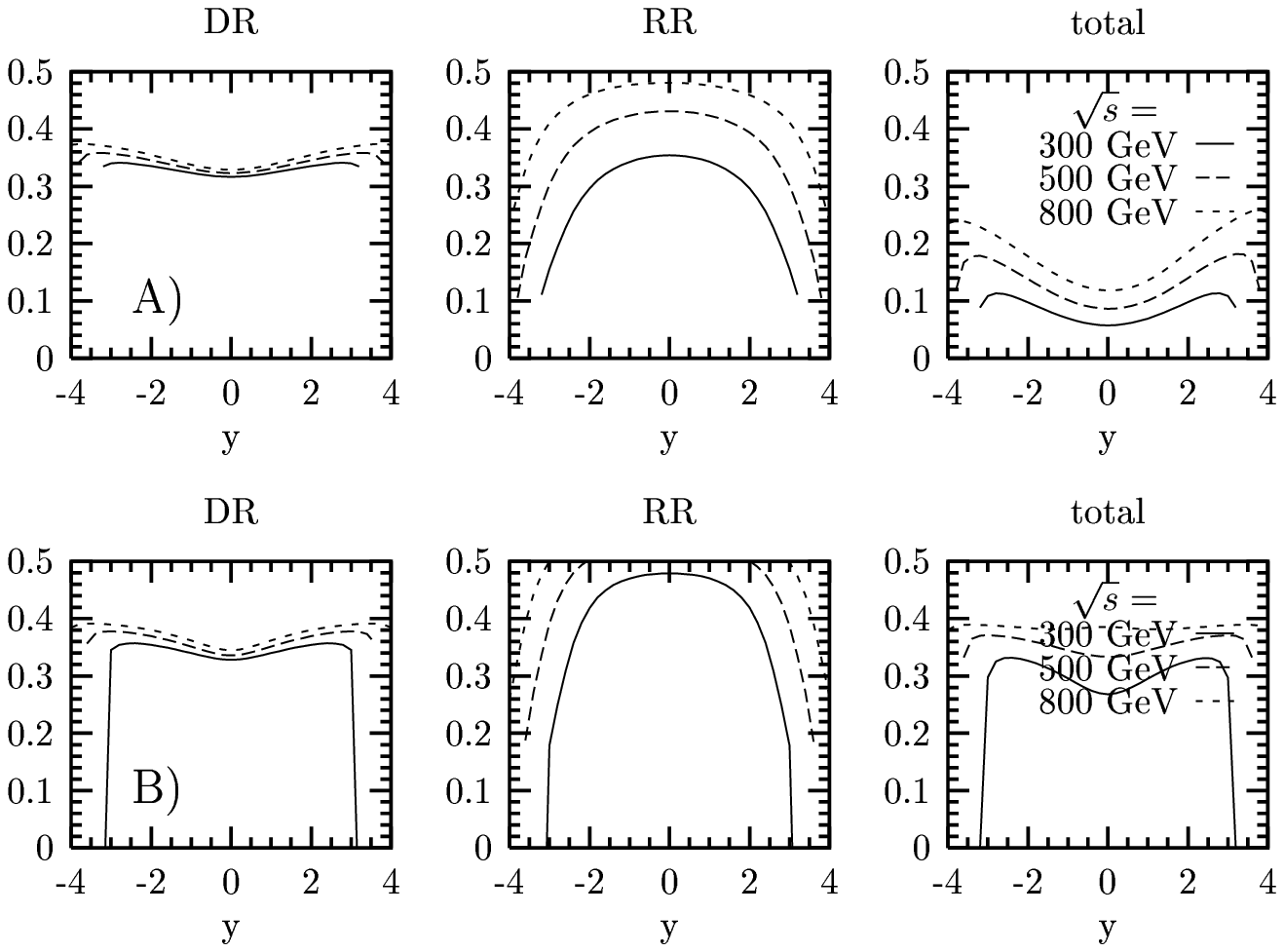}
\vskip -17cm
\caption{The ratio {\large $\frac{GRV-SAS1d}{GRV}$} of the cross-section 
{\Large $\frac{d^2\sigma}{dp^2_Tdy}$}$(e^+e^- \rightarrow e^+e^- c/\bar{c} X)$
\newline for $p_T=10$ GeV, 
in FFNS scheme, for two photon spectra: A) WW, B) LASER}
\end{figure}

\end{document}